\documentclass[acmsmall]{acmart}
\usepackage{comment}
\usepackage[section]{placeins}
\usepackage{afterpage}
\usepackage{placeins}
\usepackage{amsmath}

\AtBeginDocument{%
  \providecommand\BibTeX{{%
    \normalfont B\kern-0.5em{\scshape i\kern-0.25em b}\kern-0.8em\TeX}}}

\setcopyright{acmcopyright}
\copyrightyear{2022}
\acmYear{2022}
\acmDOI{XXXXXXX.XXXXXXX}

\acmConference[PEARC '22]{Make sure to enter the correct
  conference title from your rights confirmation emai}{July 10--14,
  2022}{Boston, MA}
\acmPrice{15.00}
\acmISBN{978-1-4503-XXXX-X/18/06}

\begin{document}

\title{Institutional Value of a Nobel Prize}

\author{Sasidev Mahendran}
\email{sasmahe@iu.edu}
\orcid{0000-0002-5124-6575}
\affiliation{%
  \institution{Pervasive Technology Institute, Indiana University}
  \streetaddress{2709 E. 10th St.}
  \city{Bloomington}
  \state{Indiana}
  \country{USA}
  \postcode{47408}
}

\author{Claudia M. Costa}
\affiliation{%
  \institution{Pervasive Technology Institute, Indiana University}
  \city{Bloomington}
  \state{Indiana}
  \country{USA}}
\email{clcosta@iu.edu}
\orcid{0000-0002-9869-1780}

\author{Julie A. Wernert}
\affiliation{
  \institution{Pervasive Technology Institute, Indiana University}
  \city{Bloomington}
  \state{Indiana}
  \country{USA}}
\email{jwernert@iu.edu}
\orcid{0000-0002-5705-9527}

\author{Craig A. Stewart}
\orcid{0000-0003-2423-9019}
\affiliation{%
  \institution{National Center for Supercomputing Applications, University of Illinois at Urbana-Champaign}
  \city{Urbana}
  \state{IL}
  \country{USA}
}
\affiliation{%
  \institution{Department of Computer Science, Indiana University}
  \city{Bloomington}
  \state{IN}
  \country{USA}
}

\renewcommand{\shortauthors}{Mahendran et al.}


\begin{abstract}
The Nobel Prize is awarded each year to individuals who have conferred the greatest benefit to humankind in Physics, Chemistry, Medicine, Economics, Literature, and Peace, and is considered by many to be the most prestigious recognition for one’s body of work. Receiving a Nobel prize confers a sense of financial independence and significant prestige, vaulting its recipients to global prominence. Apart from the prize money (approximately US\$1,145,000), a Nobel laureate can expect to benefit in a number of ways, including increased success in securing grants, wider adoption and promulgation of one's theories and ideas, increased professional and academic opportunities, and, in some cases, a measure of celebrity. A Nobel laureate's affiliated institution, by extension, also greatly benefits. Because of this, many institutions seek to employ Nobel Prize winners or individuals who have a high likelihood of winning one in the future. Many of the recent discoveries and innovations recognized with a Nobel prize were made possible only because of advanced computing capabilities. Understanding the ways in which advanced research computing facilities and services are essential in enabling new and important discoveries cannot be overlooked in examining the value of a Nobel Prize. This paper explores an institution's benefits of having a Nobel Prize winner among its ranks.
\end{abstract}


\maketitle

\section{Introduction}
Established in 1901, the Nobel Prize is one of the oldest and most elite prizes. With rare exceptions, the Nobel Prize is almost always a once-in-a-lifetime event\cite{fourNobel}. As a result, any institution that plays a part in the receipt of such an award, or has a Nobel Prize laureate among their current or past faculty, places a great amount of emphasis on this in public relations. Institutions may also invest heavily in support of researchers who are expected to win a Nobel Prize. The National Science Foundation (NSF) supported the LIGO project\cite{LIGO}, including significant funding for computing equipment, in the hope that it would lead to the verification of the existence of gravitational waves, well understood to be worthy of a Nobel Prize if achieved\cite{nsfligo}. The NSF-funded eXtreme Science and Engineering Discovery Environment (XSEDE) has played a role in three Nobel Prizes so far, and has emphasized this a great deal in its public relations materials. 


Receiving a Nobel prize confers a sense of financial independence and significant prestige, vaulting its recipients to global prominence. Apart from the prize money, a Nobel laureate can expect to benefit in a number of ways, including increased success in securing grants, wider adoption and promulgation of one's theories and ideas, increased professional and academic opportunities, and, in some cases, a measure of celebrity.  

The institution with which a Nobel laureate is affiliated, by extension, also greatly benefits. Nevertheless, it is hard to know how valuable a Nobel Prize is to institutions claiming some role in supporting the work leading to such an award. Stewart et al.\cite{stewart2019assessment} have discussed the challenges of assessing the value of qualitative outcomes of the use of advanced cyberinfrastructure resources and specifically mentioned the challenge of assessing the value of a Nobel Prize. Dozens of universities in the US boast Nobel Prize winners among its faculty; XSEDE boasts three among its community of researchers. Indeed, universities other than Indiana University have made investments in computing resources in the hopes they would result in a discovery sufficiently significant to receive a Nobel Prize. XSEDE provided computational support for research that led to two Nobel Prizes in Physics. It is often helpful to explain what the return on investment (ROI) in computer systems is or could be. Understanding the value of a Nobel Prize is a prerequisite for any attempt to understand the ROI in a cyberinfrastructure system that enables award-winning research.

With all of the attention given to Nobel Prizes, one would expect an existing way to analyze its value to institutions supporting the research. We have searched academic literature and public press and can find very little written about this topic. Our purpose in this paper is to make "first attempts" at assessing the value of the Nobel Prizes and the impact of these prizes on affiliated organizations by looking at three organizations from different sectors: General Electric, Indiana University, and the NSF-funded XSEDE. We will look at the financial impacts of Ivar Giaever’s Nobel Prize in Physics for discovering electron tunneling in superconductors, pivotal in developing General Electric's magnetic resonance imaging (MRI) technology. For the 2009 Nobel Prize in Economics awarded to Elinor Ostrom of Indiana University for her analysis of economic governance, we focus on the impacts on the institution. Finally, the 2013 Nobel Prize in Chemistry, awarded to Martin Karplus, Michael Levitt, and Arieh Warshel for their contributions in computational chemistry enabling the use of both classical and quantum physics to simulate chemical reactions, we use this case to exemplify scientific impact. XSEDE played an important role in providing the computational resources required to enable their experiments\cite{nsf}. This is very much the beginning of an approach to the challenge of assessing the value of a Nobel Prize, though certainly not a definitive conclusion. This paper, using use case methodologies, speaks more specifically to the value of a Nobel Prize to institutions supporting Nobel Prize Laureates than any other report of which we are aware.

\section{Methods}
This paper follows a case-study-based method to analyze each institution and understand the impact of a Nobel Prize on it. 
For this study, data was gathered using a variety of methods like interviews, observations, and analysis of primary and secondary sources, including articles, research papers, and official documents.

\subsection{General Electric}

In 1901, General Electric (GE) founded the Corporate Research and Development Center, the first dedicated industrial laboratory in the world. Having begun his career at GE’s Canadian division, Ivar Giaever joined the research laboratory in 1956. In 1960, he demonstrated the tunneling of electrons under superconducting conditions, which according to classical physics, had been thought impossible. He discovered how to pitch electrons through a thin layer of insulating material, earning the 1973 Nobel Prize in Physics; this work later proved pivotal to General Electric's development of the world’s first full-body MRI machine. Through this experiment, Giaever also validated the Bardeen–Cooper–Schrieffer (BCS) theory of superconductivity\cite{abetti2002science}. 

As a consequence of Giaever’s pioneering research, superconductivity seemed to have many potential applications. While GE’s business units showed an interest in the technology and expectations were high, the commitment was low, and, ultimately, a decision was made to discontinue the development of new products using this technology. Fortunately, an independent entrepreneur was interested in the technology and agreed to create a spin-off venture with General Electric, known as Intermagnetics General Corporation\cite{abetti2002science}. Intermagnetics aggressively pursued the laboratory-research-magnets business and within a year, the venture became profitable, with GE’s medical division being its major customer. With the success of this venture, GE realized that superconducting magnets were vital to the performance and reliability of MRI systems, and began to heavily invest in their design and manufacture. GE reentered the superconducting magnet business and rolled out its first full-body MRI system in 1984\cite{abetti2002science}. 

Over the course of decades, MRI systems have had a significant impact on GE’s growth. After its initial investment of \$300 million in 1979\cite{nexisuni}, the MRI division returned its first profits in 1988. GE's 1988 annual report\cite{GE1998} states that the MRI division was one of the main reasons for record levels of earnings and gains during the previous year. Furthermore, looking at the historical stock price trajectory, a significant increase in the stock price after 1988 can be clearly seen. Also, Securities and Exchange Commission filings further underscore the impact of MRI systems on GE’s increased business\cite{nexisuni}.

\begin{figure}[ht]
  \centering
  \includegraphics[width=\linewidth]{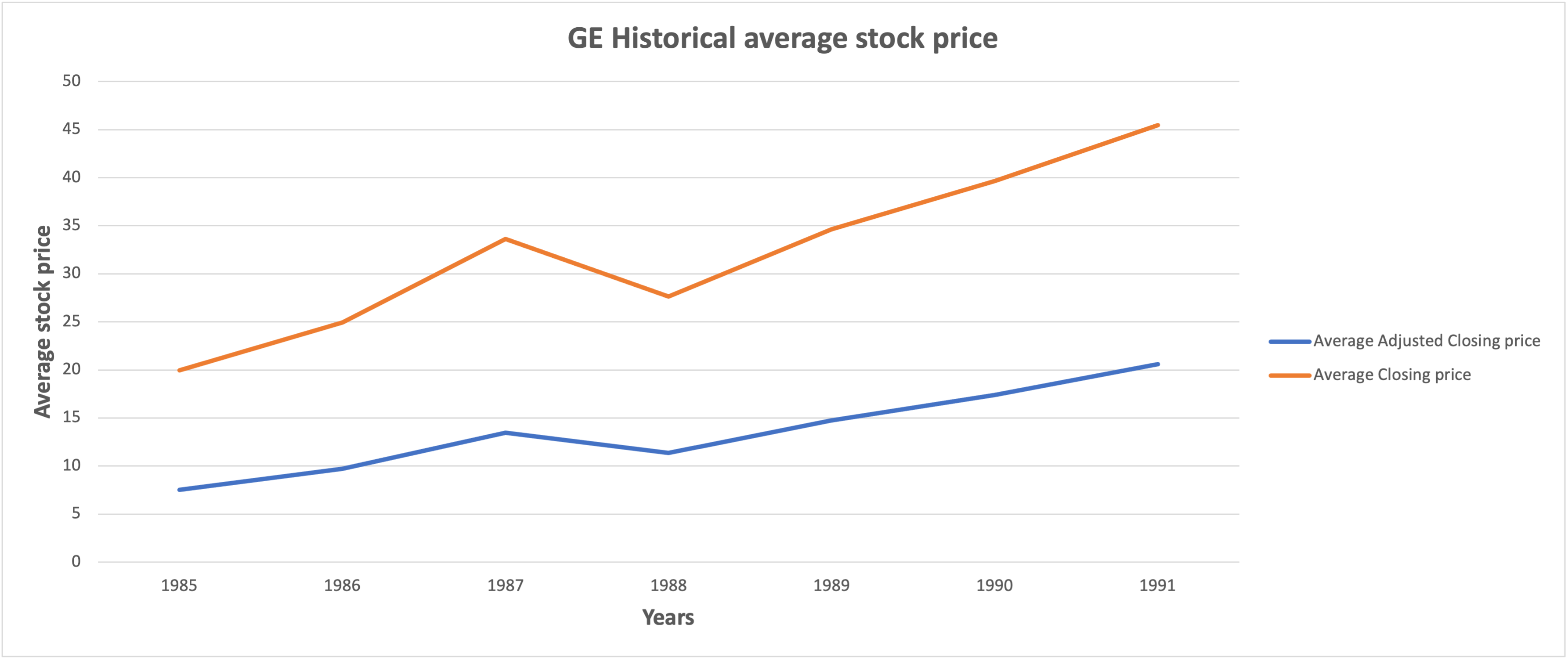}
  \caption{Historical average closing and adjusted \protect\footnotemark{} stock price every year from 1985 to 1991, via Yahoo Finance\cite{yahoo}}
  \label{fig:GE}
  \Description{-}
\end{figure}
\afterpage{\footnotetext{Average Adjusted closing price reflects any of the corporate actions upon the actual closing price}}

Looking at Fig. \ref{fig:GE}, one can see that GE’S stock price has increased steadily since 1988\cite{yahoo}, the same year the GE's MRI division showed its first profits. Thus, one can reasonably conclude that the Nobel Prize-winning research on superconductivity was pivotal in helping the company develop the MRI technology\cite{GE} and, in turn, contributed to the company’s increased profits and stock gains.

\subsection{Indiana University}

Indiana University's Distinguished Professor of Political Science, Elinor Ostrom was awarded the Nobel Prize for Economic Sciences, also known as the Sveriges Riksbank Prize, in 2009\cite{economicsNobel}. Ostrom’s work focused on economic governance, especially the cultural and natural resources accessible to all members of society (i.e., air, water, habitable land), also known as "the commons." Ostrom demonstrated the fallacy of the "tragedy of the commons" by showing how local communities can successfully manage "the commons" without any regulation by central authorities and, in turn, disproving the notion that "the commons" would inevitably be exploited and destroyed over time. This was the most recent Nobel Prize awarded to a faculty member at Indiana University\cite{economicspdf}. Michael McRobbie, who now serves as the university chancellor, was the president of Indiana University when Ostrom received her Nobel Prize. In a February 2022 interview conducted by this paper's authors, McRobbie shared his perspectives on the impacts of the Nobel Prize on the life and legacy of the institution.

During an interview conducted by the authors on February 28, 2022, Michael McRobbie, University Chancellor, Indiana University, stated that while there are dozens of other prizes that may offer more money, nothing comes close to a Nobel Prize in prestige and lasting impact. Winning a Nobel Prize obviously speaks to the person’s exemplary work. For an institution, it reflects on its reputation for academic excellence and commitment to quality. The award of a Nobel Prize to one of its faculty also speaks to the research environment present at an institution, which, in turn, allows the university to attract and retain the best and brightest faculty and students. In McRobbie’s view, Elinor Ostrom’s Nobel Prize still has a lasting impact on Indiana University. The Prize is reasserting and independently validating the excellence of IU’s research enterprise, “putting it on the map as one of a small number of elite universities whose faculty have recently won Nobel Prizes.”

Furthermore, McRobbie noted that a previously awarded Nobel Prize might influence the selection of future Nobel laureates, and “because it has an enormous impact on the perception of the institution’s academic excellence, it may also bolster the success of development campaigns and alumni donations, as well as the institution's competitiveness in winning federal research dollars.” When hiring faculty members, “the prospect of a Nobel Prize is certainly a factor not to be overlooked.” He also noted that computational resources, in which IU heavily invested during his tenure, have been crucial for Nobel Prize-winning research in recent years. 

There are nine Nobel prizes associated with IU, more than in some countries. McRobbie stressed that Ostrom’s Nobel Prize has been “galvanizing” for Indiana University, and the global recognition of her research accomplishments continues to generate tremendous goodwill. “It was a confirmation of her academic excellence, and that of Indiana University, in the strongest possible way.”

\subsection{XSEDE (The eXtreme Science and Engineering Discovery Environment)}

The 2013 Nobel Prize in Chemistry was awarded to Arieh Warshel, Martin Karplus, and Michael Levitt for the development of multi-scale models for complex chemical systems. They developed a method to make Newton’s classical physics work side-by-side with the fundamentally different quantum physics\cite{chemistryNobel}. Classical physics is easily used to model large molecules, but it is impossible to simulate chemical reactions. Conversely, quantum physics is used to simulate chemical reactions, but it requires enormous amounts of computing power, as the computer has to process every single electron and nucleus in a molecule\cite{chemistrypdf}. As recently as the early 1970s, scientists were only able to simulate the chemical reactions of small molecules. Karplus and his research group had developed computer programs that could simulate chemical reactions with the help of quantum physics\cite{chemistrypdf}. 

Conversely, Warshel and Levitt developed a computer program based on classical theories to model all kinds of molecules. Eventually, Karplus and Warshel joined together to develop a computer program that drew on quantum physics and applied classical theories. A computer program enabling the simulation of molecular dynamics, called CHARMM, was first developed by Karplus’s research group at Harvard in the late 1960s\cite{brooks2009charmm}. CHARMM was further improved by Arieh Warshel’s consistent force field program and Michael Levitt’s pioneering energy calculations for proteins. CHARMM is one of the key software products used by researchers and scientists all over the world for generating and analyzing a wide range of molecular simulations. 

\begin{figure}[h]
  \centering
  \includegraphics[width=\linewidth]{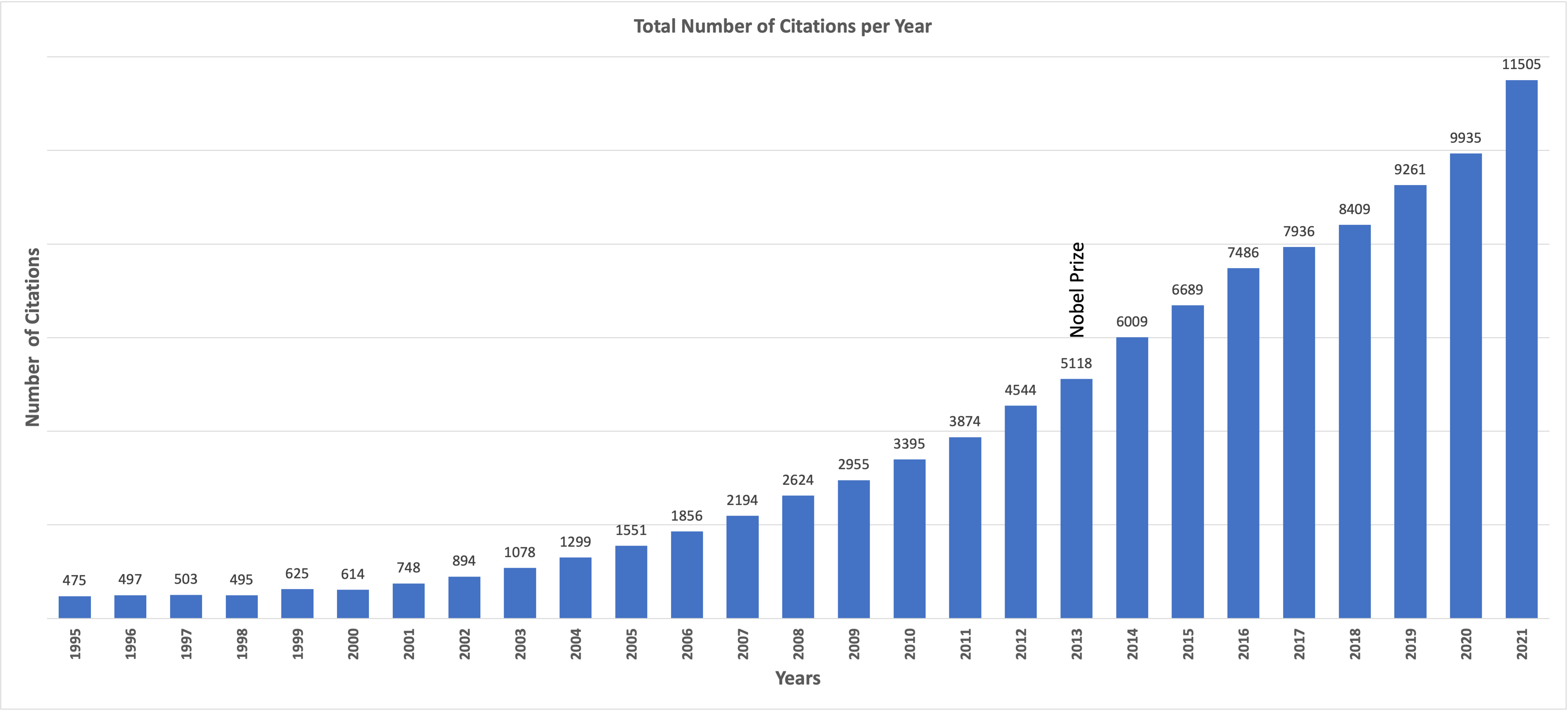}
  \caption{Number of times CHARMM was cited every year from 1995 to 2021, via Clarivate Web of Science.\cite{webofscience}}
  \label{fig:CHARMM}

  \Description{-}
\end{figure}

Karplus, Levitt, and Warshel have all used National Science Foundation (NSF)-supported cyberinfrastructure to advance their research, namely TeraGrid and its successor, XSEDE\cite{bringhpctolab}. Both are advanced cyberinfrastructure systems that have enabled researchers to access and share high performance computing resources. Karplus used supercomputing resources, including the Queenbee, Big Ben, and Lonestar supercomputers that were part of TeraGrid\cite{nsf}. Though XSEDE is not directly connected to the 2013 Chemistry Nobel Prize, it played a pivotal role in developing and popularizing the CHARMM software. CHARMM has had a significant impact on the scientific community and is widely used by researchers across the world. To better understand its impact, a scientific impact analysis of the CHARMM software is taken from the Web of Science\cite{webofscience}, as shown in Fig. \ref{fig:CHARMM}.

\section{Conclusion}

The Nobel Prize remains the most prestigious and respected award for scientific research. Winning a Nobel prize creates lasting, multi-fold impacts for the individual and the affiliated organizations. As seen from the above case studies, the value of a Nobel Prize cannot be simply or practically connected to a number or reduced to a calculation. 

For General Electric, the Nobel Prize-winning research directly contributed to the development of modern MRI technology, thereby having vast financial impact – not to mention saving countless lives – over the course of decades which continue today. Elinor Ostrom’s Nobel Prize created a considerable impact in terms of goodwill and prestige for Indiana University, resulting in increased federal and philanthropic funding and the attraction and retention of the finest faculty and students. Finally, the CHARMM software, one of the results of research by Karplus, Warshel, and Levitt, continues to have significant impacts on the global scientific community.

This paper by no means can or should be the last word on assessing the overall value or impact of a Nobel Prize on an organization that supports research leading to discoveries acknowledged with this prize. Nonetheless, this paper makes clear that this is an interesting question which requires further consideration and research. While there is no clear and simple answer to this question, we have shown that use case analyses are one way to productively navigate toward an answer. 


\begin{acks}
This work was supported in part by NSF award 1548562 in support of XSEDE2 (John Towns, PI),  by support from the IU Pervasive Technology Institute and by the UITS Bepko Internship program. Thanks to Chancellor Michael McRobbie for his support and contribution, which helped us to  fully understand the impacts of Elinor Ostrum's 2009 Nobel Prize in Economics on Indiana University. Thanks to Marlon Pierce for his valuable comments and feedback. Thanks to Kristol Hancock and Tonya Miles for editing. Any opinions presented here are those of the authors and may not reflect those of supporting organizations.

\end{acks}

\bibliographystyle{ACM-Reference-Format}
\bibliography{citation}
\end{document}